\documentclass[
11pt,onecolumn
]{article}

\usepackage[T1]{fontenc}
\usepackage[utf8]{inputenc}
\usepackage{lmodern}

\usepackage{amsmath}
\usepackage{amssymb,comment}
\usepackage{amsfonts}
\usepackage{mathtools}

\usepackage{graphicx}
\usepackage{caption}

\usepackage[
  margin=1.65cm,
  columnsep=0.65cm
]{geometry}

\usepackage{cite}
\usepackage[hidelinks]{hyperref}

\numberwithin{equation}{section}

\title{Inverse characterization and non-uniqueness of the Gaussian configurational partition function}

\author{
    Ignacio S. Gomez$^{1,2}$
    \\[3pt]
    $^{1}$Departamento de Ciências Exatas e Naturais, Universidade Estadual do Sudoeste da Bahia, \\
			    BR 415, Itapetinga - BA, 45700-000, Brazil
    \\[3pt]
    $^{2}$PROFÍSICA – Programa de Pós-Graduação em Física, 
Universidade Estadual de Santa Cruz,\\ 45650-000 Ilhéus, BA, Brazil
    \\[3pt]
    \small
    \texttt{ignacio.gomez@uesb.edu.br}
}

\date{}

\begin{document}

\maketitle

\begin{abstract}
Given a configurational partition function, in this work we investigate 
the inverse problem of reconstructing the one-dimensional
potential. 
For the case of the Gaussian partition function, the configurational density of states (CDOS) is univocally
determined, being the harmonic potential uniquely recovered due to
its symmetric single--well feature. 
When multiples inverse branches are considered, the uniqueness of the potential is broken and more information is needed in order to obtain a unique potential. 
The branch topology imposes alternating orientations, thus
allowing distinct spatial realizations. Finally, the Gaussian partition
function only determines the CDOS but not the potential, manifesting in this way the precise scope and limitations of the inverse characterization of the Gaussian configurational partition function.
\end{abstract}

\vspace{0.1cm}


\section{Introduction}
\label{sec:introduction}

The configurational partition function is a cornerstone quantity in
classical statistical mechanics, which for a one-dimensional system under a
potential $V(x)$ is given by
\begin{equation}
Z(\beta)
=
\int_{\mathbb R} e^{-\beta V(x)}\,dx,
\qquad
\beta=(k_{\mathrm B}T)^{-1},
\label{eq:Z_intro}
\end{equation}
and the determination of $Z(\beta)$ from $V(x)$ represents the direct problem. In this work we consider the inverse problem: In what way can
the potential be reconstructed from its configurational partition
function? This question 
led to fundamental concerns in the classical formulation of
statistical mechanics, by means of the kinetic and probabilistic frameworks
of Maxwell and Boltzmann to the thermodynamic formalism of Gibbs and
its subsequent mathematical developments
\cite{Maxwell,Boltzmann,Gibbs,Tolman,Khinchin,Pathria,Ruelle,Georgii,Kirsch,Isakov}. A natural answer of the inverse problem can be expressed by introducing
the configurational density of states (CDOS),
\begin{equation}
\rho_V(u)
=
\int_{\mathbb R}
dx\,\delta\!\left(u-V(x)\right),
\label{eq:rho_intro}
\end{equation}
where the partition function is written as
\begin{equation}
Z(\beta)
=
\int_{V_{\min}}^\infty
\rho_V(u)e^{-\beta u}\,du,
\label{eq:laplace_intro}
\end{equation}
establishing that the configurational partition function is simply the Laplace
transform of the density of states with the corresponding inverse
transform being a standard problem in the Laplace transform theory
\cite{Widder}.
In the microcanonical ensemble, the CDOS plays a key role in establishing the the statistical description of
finite systems \cite{Naudts}.
Recent developments of the CDOS have been made concerning simple solids \cite{Davis2023}, reconstruction of thermodynamics from statistical data 
\cite{Moreno2023}, connections between position-dependent systems and superestatistics \cite{SantosGomezDaCostaMustafa2021,GomezSantosSantosThibes2025}, among others. This increasing interest of the 
partition function and the DOS
as carriers of information of the underlying energy scenario motivated the main objective of the present work: the reconstruction of the potential from a given the configurational partition function. 
For an energy value $u$ let  $x_i(u)$ denote the 
solutions of
\begin{equation}
V(x_i)=u.
\end{equation}
From the property of the Dirac delta
distribution Eq.~\eqref{eq:rho_intro} is recasted as
\begin{equation}
\rho_V(u)
=
\sum_i\frac{1}{|V'(x_i)|}
=
\sum_i\left|\frac{dx_i}{du}\right|.
\label{eq:branch_density_intro},
\end{equation}
thus showing that the DOS contains the contribution of all
inverse branches. The deduction of the potential $V(x)$ from Eq. \eqref{eq:branch_density_intro} is thus an inverse problem 
where the 
multiplicity and the geometry of the level sets play a crucial role
\cite{Kirsch,Isakov}, with the particularity that Eq.~\eqref{eq:branch_density_intro} contains only the
sum of the absolute branch slopes and from this follows that any CDOS can not determine the spatial dependence of the potential univocally. 

The goal of this work is to establish a definite scope of the inverse problem for the Gaussian configurational partition function \eqref{eq:Z_intro} given by the harmonic potential $V(x)=\frac{k}{2}x^2$. We first recover the harmonic potential within the
symmetric two-branch class and  then we define the suitable four-branch
construction and demonstrate that the Gaussian CDOS restrains
the possible branch slopes, leaving freedom in their individual realizations. 
This non-uniqueness 
determines the difference between 
the complete recovery of the potential and its generally non-unique recovery.    
The paper is organized as follows. Section~\ref{sec:gaussian_inverse}
develops the Gaussian inverse characterization of the harmonic potential. Section
\ref{sec:multibranch} establishes the multi-branch non-uniqueness.
Section~\ref{sec:discussion} discusses the consequences, and
Section~\ref{sec:conclusions} summarizes the results.

\section{Gaussian inverse characterization}
\label{sec:gaussian_inverse}

\begin{figure}[t]
\centering
\includegraphics[width=\columnwidth]{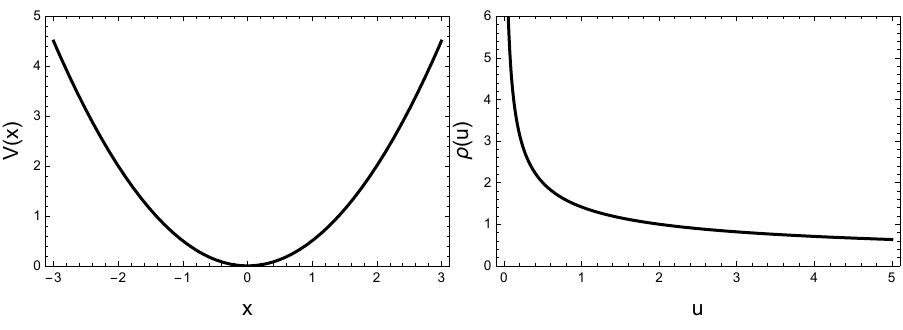}
\caption{Gaussian configurational case. (a) Harmonic potential
$V_{\rm G}(x)=kx^2/2$. (b) CDOS
$\rho_{\rm G}(u)=\sqrt{2/k}\,u^{-1/2}$ deduced from the inverse
Laplace transform of
$Z_{\rm G}(\beta)=\sqrt{2\pi/(\beta k)}$.}
\label{fig:gaussian_reference}
\end{figure}

\begin{figure}[t]
\centering
\includegraphics[width=\columnwidth]{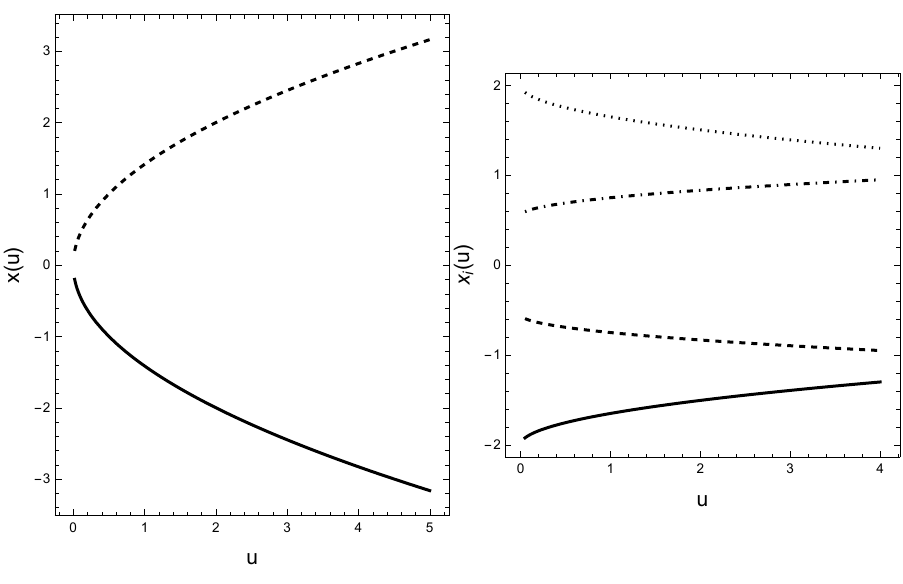}
\caption{Inverse-branch structure. (a) Gaussian two-branch
reconstruction $x_\pm(u)=\pm\sqrt{2u/k}$, with
$x_-'(u)<0$ and $x_+'(u)>0$. (b) Four-branch structure
$x_1<x_2<x_3<x_4$, showing the alternating orientation
$x_1'<0$, $x_2'>0$, $x_3'<0$, and $x_4'>0$.}
\label{fig:inverse_branches}
\end{figure}

Let
\begin{equation}
Z_{\rm G}(\beta)
=\int_{\mathbb R} e^{-\beta \frac{k}{2}x^2}\,dx=
\sqrt{\frac{2\pi}{\beta k}},
\qquad
k>0,
\label{eq:Z_G}
\end{equation}
be the Gaussian configurational partition function.
From Eq.~\eqref{eq:laplace_intro} and inverting the Laplace transform, the CDOS of \eqref{eq:Z_G} results 
\begin{equation}
\rho_{\rm G}(u)
=
\mathcal{L}^{-1}_{\beta\rightarrow u}
\left[
Z_{\rm G}(\beta)
\right].
\end{equation}
Using
\begin{equation}
\mathcal{L}
\left\{
u^{-1/2}
\right\}
=
\sqrt{\pi}\,\beta^{-1/2},
\end{equation}
we obtain
\begin{equation}
\boxed{
\rho_{\rm G}(u)
=
\sqrt{\frac{2}{k}}\,u^{-1/2},
\qquad
u>0
}
\label{eq:rho_G}
\end{equation}
Thus, we see that the Gaussian dependence of the partition function determines univocally 
the CDOS. 
The Gaussian reference potential and its DOS are shown
in Fig.~\ref{fig:gaussian_reference},
where every regular energy level
above the minimum has two inverse branches,
\begin{equation}
x_-(u)=-f(u),
\qquad
x_+(u)=f(u),
\end{equation}
with $f'(u)>0$. By replacing these branches in Equation~\eqref{eq:branch_density_intro} results 
\begin{equation}
\rho_V(u)=2f'(u).
\label{eq:rho_two_branch}
\end{equation}
From this expression and Eq.~\eqref{eq:rho_G} it follows
\begin{equation}
f'(u)
=
\frac{1}{\sqrt{2ku}},
\label{eq:fprime_gaussian}
\end{equation}
whose integration yields
\begin{equation}
f(u)
=
\sqrt{\frac{2u}{k}}+C,
\label{eq:f_gaussian}
\end{equation}
with $C$ a spatial translation constant. If we choose the origin at
the minimum we have $C=0$, and thus
\begin{equation}
x_\pm(u)
=
\pm\sqrt{\frac{2u}{k}}.
\label{eq:inverse_gaussian_branches}
\end{equation}
The inverse branches 
are illustrated in
Fig.~\ref{fig:inverse_branches}.
By inverting Eq.~\eqref{eq:inverse_gaussian_branches} we have
\begin{equation}
u=\frac{kx^2}{2},
\end{equation}
and then
\begin{equation}
\boxed{
V(x)=\frac{kx^2}{2}+V_0.
}
\label{eq:reconstructed_harmonic}
\end{equation}
This establishes an inverse characterization of the
Gaussian configurational partition function within the class of
symmetric single-well potentials on the real line, where the Gaussian
partition function determines uniquely the harmonic potential except for a spatial translation $C$ and an energy constant $V_0$.

The uniqueness is a consequence of the two-branch structure and the reflection symmetry of the harmonic potential.
Without these constraints, Eq.~\eqref{eq:rho_G} assigns only the
total measure of the inverse branches,
\begin{equation}
\rho_{\rm G}(u)
=
\sum_i
\left|
\frac{dx_i}{du}
\right|,
\label{eq:aggregate_constraint}
\end{equation}
leaving the individual functions $x_i(u)$ undetermined.

\section{Multi-branch case and non-uniqueness}
\label{sec:multibranch}

\begin{figure}[t]
\centering
\includegraphics[width=\columnwidth]{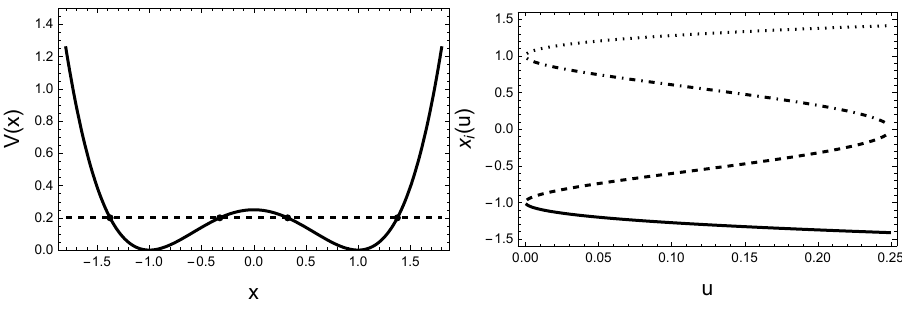}
\caption{Four-branch topology of a symmetric double-well potential.
(a) The quartic potential
$V_{\rm DW}(x)=\lambda(x^2-a^2)^2/4$. The horizontal level corresponds
to an energy below the central barrier. (b) The four inverse branches
associated with this energy range exhibiting the required
orientation $(-,+,-,+)$. Above the barrier, the two inner branches
cancel. The double-well is used only to illustrate the topology of
the inverse branches and is not claimed to reproduce the Gaussian
partition function.}
\label{fig:double_well}
\end{figure}

Let consider the inverse problem when an energy level has
more than two spatial preimages. Consider $V(x)$ a smooth
potential on the real line that satisfies
\begin{equation}
V(x)\longrightarrow+\infty
\qquad
\text{as}
\qquad
x\longrightarrow\pm\infty.
\label{eq:proper_potential}
\end{equation}
Assume that a value $u$ has four ordered simple solutions
\begin{equation}
x_1(u)<x_2(u)<x_3(u)<x_4(u),
\qquad
V(x_i(u))=u.
\label{eq:four_roots}
\end{equation}
Because $V(x)-u$ is positive at $x=\pm\infty$ at every simple crossing the sign of $V(x)$ alternates. Hence,
\begin{equation}
V'(x_1)<0 \ , \
V'(x_2)>0 \ , \
V'(x_3)<0 \ , \
V'(x_4)>0,
\label{eq:four_Vprime_signs}
\end{equation}
and then
\begin{equation}
\boxed{
x_1'<0,\qquad
x_2'>0,\qquad
x_3'<0,\qquad
x_4'>0.
}
\label{eq:four_branch_signs}
\end{equation}
For a symmetric four-branch case we can write
\begin{equation}
x_{1,4}(u)=\mp f(u) \quad , \quad 
x_{2,3}(u)=\mp g(u),
\label{eq:symmetric_four_branches}
\end{equation}
so by Eq. \eqref{eq:four_branch_signs} it follows
\begin{equation}
f'(u)>0,
\qquad
g'(u)<0.
\label{eq:fg_signs}
\end{equation}
This indicates that the outer branches move outward as the energy increases while the inner branches move in the opposite direction.
The corresponding DOS is
\begin{equation}
\rho_V(u)=
\sum_{i=1}^{4}
\left|x_i'(u)\right|
\nonumber=
2f'(u)-2g'(u).
\label{eq:rho_four_branch}
\end{equation}
By imposing the Gaussian density of states we obtain
\begin{equation}
2\left[f'(u)-g'(u)\right]
=
\sqrt{\frac{2}{k}}\,u^{-1/2},
\label{eq:gaussian_four_constraint}
\end{equation}
from which after integration we have
\begin{equation}
f(u)-g(u)
=
\sqrt{\frac{2u}{k}}+C.
\label{eq:fg_difference}
\end{equation}
Differently from the two-branch case, Eq~\eqref{eq:fg_difference} is not capable to
determine $f$ and $g$ separately but only fixes their difference.
Suppose $g(u)$ an admissible inner branch satisfying
\begin{equation}
g(u)>0,
\qquad
g'(u)<0,
\end{equation}
and let define
\begin{equation}
f'(u)
=
g'(u)+
\frac{1}{\sqrt{2ku}}.
\label{eq:f_from_g}
\end{equation}
Any time
\begin{equation}
f'(u)>0,
\label{eq:f_admissibility}
\end{equation}
the four branches posses the correct orientation and reproduce the
Gaussian DOS. In this way, we see that different choices of $g(u)$ can lead
to distinct spatial realizations having the same configurational
partition function. It is worth mentioning that 
the distinction between branch orientation and the DOS
constraint is essential. For instance, a parametrization of the form
\begin{equation}
x_1=-f,\qquad
x_2=-g,\qquad
x_3=g,\qquad
x_4=f
\end{equation}
being $f$ and $g$ increasing leads to
\begin{equation}
(x_1',x_2',x_3',x_4')
=
(-,-,+,+),
\end{equation}
that results incompatible with Eq.~\eqref{eq:four_branch_signs}.
A symmetric double-well provides a useful 
illustration of this.
Consider
\begin{equation}
V_{\rm DW}(x)
=
\frac{\lambda}{4}(x^2-a^2)^2.
\end{equation}
For energies below the central barrier,
\begin{equation}
0<u<\frac{\lambda a^4}{4},
\end{equation}
so we have four spatial intersections. Above the barrier, the two
inner branches cancel and only the two outer branches are preserved.
This type of topology changes are the relevant structural information in an inverse problem
\cite{Kirsch,Isakov}.
The double-well topology and its functional freedom are illustrated in
Fig.~\ref{fig:double_well} and Fig.~\ref{fig:inverse_branches} respectively.
\begin{figure}[t]
\centering
\includegraphics[width=\columnwidth]{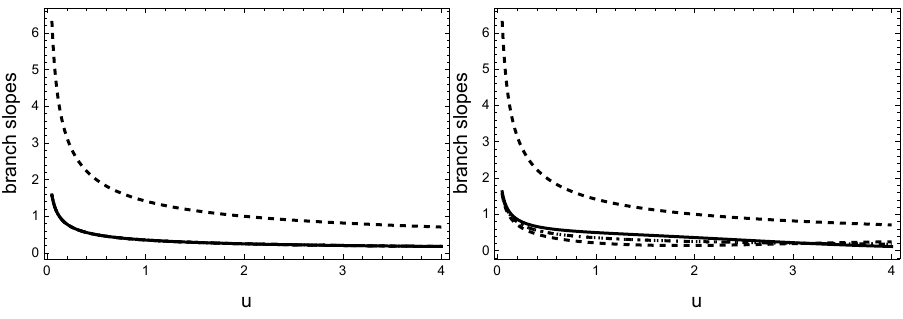}
\caption{Two different decompositions of the Gaussian CDOS into inverse-branch contributions. In both cases
the sum of the four positive branch slopes is
$\rho_G(u)=\sqrt{2/u}$, while the individual contributions are distinct,
thus illustrating the non-uniqueness of the multi-branch inverse
construction.}
\label{fig:nonuniqueness}
\end{figure}
Thus, the multi-branch analysis leads to the central result
\begin{equation}
\boxed{
Z_{\rm G}(\beta)
\Longrightarrow 
\rho_{\rm G}(u)
\quad\text{uniquely} \ , \
\rho_{\rm G}(u)
\not\Longrightarrow 
V(x)
\quad\text{uniquely}.
}
\label{eq:central_inverse_result}
\end{equation}
The harmonic potential is recovered uniquely only 
for the class which has the two-branch
structure and symmetry.

\section{Consequences and discussion}
\label{sec:discussion}

The direct construction
\begin{equation}
V(x)
\longrightarrow
\rho_V(u)
\longrightarrow
Z(\beta)
\end{equation}
encompasses two different mappings. The first provides a CDOS given an arbitrary potential, while the second is the Laplace
transform in Eq.~\eqref{eq:laplace_intro}.
The inverse problem says that these mappings are not equivalent since the inverse transform only determines the DOS, leaving the reconstruction of the potential subjected to the geometry and multiplicity of its level sets
\cite{Widder,Kirsch,Isakov}.
The Gaussian case shows 
that its symmetric single-well implies the density
\begin{equation}
\rho_{\rm G}(u)
=
\sqrt{\frac{2}{k}}\,u^{-1/2}
\end{equation}
which fixes the two inverse branches and thus the harmonic
potential. The uniqueness results a consequence not only of the
Gaussian partition function but also of the constraints. 
For the case of several branches, the same density restrains only
the quantity
\begin{equation}
\sum_i
\left|
\frac{dx_i}{du}
\right|.
\end{equation}
which for the symmetric four-branch construction results
\begin{equation}
2(f'-g')
=
\sqrt{\frac{2}{k}}\,u^{-1/2},
\end{equation}
thus allowing functional freedom in the individual branches so supplementary
information is needed to restore uniqueness like 
number of wells, symmetry, energies extrema locations, or other properties of the
potential.
From a measure-theoretic outlook, the CDOS is the pushforward of
the spatial measure under the mapping $x\mapsto V(x)$. Therefore, the density
$\rho_V(u)$ registers how much spatial measure is linked
with each energy value, but not how this is distributed among
the different spatial branches. This 
characterization is intrinsically established using measure theory and level sets
\cite{Federer,EvansGariepy,Bogachev}.
Furthermore, this result provides an equivalence class of potentials.
Let $V_1,V_2$ be two potentials satisfying
\begin{equation}
\rho_{V_1}(u)=\rho_{V_2}(u),
\label{eq:equal_density}
\end{equation}
then their configurational partition functions are equal,
\begin{equation}
Z_{V_1}(\beta)=Z_{V_2}(\beta).
\end{equation}
Accordingly, the partition function characterizes an equivalence
class of potentials whose single classes correspond to constraints in such a way the spatial realization is unique. This fact is
consistent with the statistical-mechanical interpretation of
thermodynamic information in terms of densities of states
\cite{Boltzmann,Gibbs,Pathria,DavisMaulen}.
Since the two inverse problems
\begin{equation}
Z(\beta)\longrightarrow\rho_V(u)
\end{equation}
and
\begin{equation}
\rho_V(u)\longrightarrow V(x)
\end{equation}
are mathematically different then the non-uniqueness does not contradict the uniqueness
of the inverse Laplace transform. The former is governed by the uniqueness
properties of the Laplace transform \cite{Widder} while the later embraces
the multiplicity and geometry of the level sets of $V$.
Consequently, 
when multiple inverse
branches are present, the same thermodynamic information can
correspond to different spatial realizations. Thus, the inverse
characterization is unique only at the level of the
CDOS, but not generally at the level of
the potential.

\section{Conclusions}
\label{sec:conclusions}

We have presented an study of the 
inverse problem of reconstructing a
one-dimensional potential from its configurational partition function.
Our main result lies in that the partition function uniquely determines
the CDOS by means of the inverse Laplace
transform, but not the underlying potential in an univocal way.


For the Gaussian partition function case and given supplementary information about symmetric single-well, the CDOS implies a unique harmonic potential, thus expressing the uniqueness as a consequence of the Gaussian
partition function together with the two-branch structure.

When multiple inverse branches are considered, 
for a potential with four simple intersections of energy level the orientations necessarily alternate.
This can be explained since the Gaussian density $\propto u^{-1/2}$ only restrains the sum of the moduli of the branch slopes, leaving freedom for their spatial realizations given by the potential. 

Finally, our analysis allows to separate two levels of inverse
characterization: the unique recovery of the CDOS and the generally non-unique recovery of the potential.
Extra information about the potential is needed in order to obtain a unique spatial realization of the potential.
The present work provides a characterization of the scope and limitations of the inverse reconstruction of the potential
from configurational partition functions, and we hope to investigate in more dimensions in future researches.

\section*{Acknowledgements}

Ignacio S. Gomez acknowledges support from the Department of Exact
and Natural Sciences of the State University of Southwest
Bahia (UESB), Itapetinga, Bahia, Brazil, from the PROFÍSICA (UESC), Ilhéus, Bahia, Brazil and from the Conselho Nacional de Desenvolvimento Científico e Tecnológico (CNPq), Grant Number
316131/2023-7.

\end{document}